\documentclass[onecolumn,showpacs,preprintnumbers,amsmath,amssymb,aps,prd,nofootinbib,superscriptaddress]{revtex4-2}
\usepackage{graphics}
\usepackage{bm}
\usepackage{mathrsfs}
\usepackage{latexsym,amssymb,amsmath,amsfonts} 
\usepackage[english]{babel} 
\usepackage[OT1]{fontenc}
\usepackage[latin1]{inputenc}
\usepackage{makeidx}
\usepackage{hyperref}
\usepackage{color,xcolor,graphics,wrapfig,epsf}
\usepackage{epsfig,subfigure}
\usepackage[all]{xy}
\usepackage{slashed}
\DeclareGraphicsExtensions{.pdf,.jpeg,.png,.jpg}

\usepackage{comment}
\usepackage[colorinlistoftodos]{todonotes} 

\definecolor{red}{rgb}{1,0,0}

\def\p{\partial}
\def\wt{\widetilde}

\def\+{^\dagger}

\def\<{\leftarrow}
\def\>{\rightarrow}

\def\({\left(}
\def\){\right)}

\def\dual{{\,{}^\star\!}}
\def\ve{\varepsilon}
\def\a{\alpha} \def\b{\beta} \def\g{\gamma} \def\d{\delta} \def\e{\epsilon}\def\ve{\varepsilon}
\def\m{\mu} \def\n{\nu} \def\r{\rho}  \def\l{\lambda} \def\th{\theta}
\def\k{\kappa}\def\G{\Gamma}

\def\q{\quad}\def\qq{\qquad}

\def\FF{{\cal F}}\def\GG{{\cal G}}

\newcommand{\bi}{\begin{itemize}}           \newcommand{\ei}{\end{itemize}}
\newcommand{\benu}{\begin{enumerate}}   \newcommand{\enu}{\end{enumerate}}
\newcommand{\bd}{\begin{dinglist}{0}}     \newcommand{\ed}{\end{dinglist}}
\newcommand{\bfig}{\begin{figure}[htbp]}  \newcommand{\efig}{\end{figure}}
       
\newcommand{\bc}{\begin{center}}                  \newcommand{\ec}{\end{center}}
\newcommand{\be}{\begin{equation}}              \newcommand{\ee}{\end{equation}}
\newcommand{\bsub}{\begin{subequations}}  \newcommand{\esub}{\end{subequations}}
\newcommand{\ben}{\begin{eqnarray}}             \newcommand{\een}{\end{eqnarray}}
\newcommand{\ba}[1]{\begin{array}{#1}}      \newcommand{\ea}{\end{array}}
\newcommand{\bea}{\begin{equation}\begin{array}{rcl}}
\newcommand{\eea}{\end{array}\end{equation}}

\newcommand{\becomm}{}

\begin{document}
\title{Determinantal Born-Infeld Coupling of Gravity and Electromagnetism}

\author{Victor I. Afonso} \email{viafonso@df.ufcg.edu.br}
\affiliation{Unidade Acad\^{e}mica de F\'{\i}sica, Universidade Federal de Campina
    Grande, 58.429-900 Campina Grande-PB, Brazil}
\affiliation{Departamento de F\'{\i}sica, Universidad del Pa\'{\i}s Vasco
UPV/EHU, Leioa-48940, Spain}

\author{Cecilia Bejarano} \email{cbejarano@iafe.uba.ar}
\affiliation{Instituto de
Astronom\'\i a y F\'\i sica del Espacio (IAFE, CONICET-UBA), Casilla
de Correo 67, Sucursal 28, 1428 Buenos Aires, Argentina.}

\author{Rafael Ferraro} \email{ferraro@iafe.uba.ar}
\affiliation{Instituto de
Astronom\'\i a y F\'\i sica del Espacio (IAFE, CONICET-UBA), Casilla
de Correo 67, Sucursal 28, 1428 Buenos Aires, Argentina.}
\affiliation{Departamento de F\'\i sica, Facultad de Ciencias
Exactas y Naturales, Universidad de Buenos Aires, Ciudad
Universitaria, Pabell\'on I, 1428 Buenos Aires,
Argentina.}

\author{Gonzalo J. Olmo} \email{gonzalo.olmo@uv.es}
\affiliation{Departamento de F\'{i}sica Te\'{o}rica and IFIC, Centro
Mixto Universidad de Valencia - CSIC. Universidad de Valencia,
Burjassot-46100, Valencia, Spain}

\date{\today}

\begin{abstract}
We study a Born-Infeld inspired model of gravity and
electromagnetism in which both types of fields are treated on an
equal footing via a determinantal approach in a metric-affine
formulation. Though this formulation is {\it a priori} in conflict
with the postulates of metric theories of gravity, we find that the
resulting equations {can also be obtained from an action combining
 the Einstein-Hilbert action with a minimally coupled
nonlinear electrodynamics. As an example, the dynamics is solved for
the charged static black hole.}
\end{abstract}

\maketitle

\section{Introduction}
The study of alternative theories of gravity and matter that may
allow to provide a more satisfactory description of Nature has
experienced a boost in the last two decades or so, though several
examples have been known from much earlier. An important such
example is the Born-Infeld theory of electrodynamics
\cite{Born:1934gh}, in which the Maxwell Lagrangian is classically
modified {to bound the electric field intensity, so} curing
the electron self-energy problem of classical electrodynamics.
Boosted by the seminal work by Fradkin and Tseytlin
\cite{Fradkin:1985qd}, determinantal actions also found relevant
applications in M-theory scenarios to describe charged D-branes.
Following the philosophy of the Born-Infeld electromagnetic theory,
attempts to improve the gravitational dynamics at high curvatures
have also made Born-Infeld inspired proposals very attractive in
more recent years. In particular, Deser and Gibbons
\cite{Deser:1998rj} considered a theory based on the determinant of
the sum of the space-time metric plus the Ricci tensor, but it
turned out to be plagued by ghost-like instabilities. A
reformulation of this theory in the metric-affine framework was then
proposed by Vollick \cite{Vollick:2003qp}, showing that the ghosts
are removed. This new model was popularized by Ba\~{n}ados and Ferreira
\cite{Banados:2010ix}, who found that it could also help avoid
cosmological singularities. Numerous applications in cosmology,
astrophysics, and many other scenarios followed those works, and the
reader is referred to the review article
\cite{BeltranJimenez:2017doy} for a detailed account of the related
literature.

The majority of the Born-Infeld inspired modifications of
gravitational theories tend to include the matter following the
standard minimal coupling prescription of metric theories of gravity
(MTGs) \cite{Will:1993}, which is a practical rule to make the
theory compatible with the Einstein equivalence principle
\cite{Will:2014kxa}. This rule simply splits the total action into a
gravitational part plus a matter part,
$S_{Total}=S_{Gravity}+S_{Matter}$, the latter being constructed
from the Minkowskian theory by promoting the flat metric
$\eta_{\mu\nu}$ to a curved space-time metric $g_{\mu\nu}$, using
the language of differential geometry and some minimal coupling
prescription (which is not always free of ambiguities
\cite{Delhom:2020hkb}), {\it i.e.},
$S_{Matter}=S[g_{\mu\nu},\Psi_m,\nabla \Psi_m]$, with $\Psi_m$
generically representing the matter fields.

Perhaps for this reason, attempts to improve simultaneously the
matter and gravitational sectors by combining both in a single
determinantal-type action have only received timid attention, with
the notable exception, to our knowledge, of the models considered by
Vollick in \cite{Vollick:2005gc}, who proposed an action of the form
\begin{equation}\label{actionM}
I \propto \int dx^4 \left[\sqrt{-\det\(g_{\m\n}+
\e {R_{\m\n}}(\G) +\b M_{\m\n} \)}  - \l \sqrt{-g}\right] \ ,
\end{equation}
where $M_{\m\n}$ represents a quantity constructed with the
matter fields, $\e$ and $\b$ are suitable coupling constants, and
$\lambda$ determines the asymptotic behaviour of the vacuum
solutions, with $\lambda=1$ yielding Minkowski space-time. For
(massless) scalar fields one can take $M_{\m\n}=\partial_\mu\phi
\partial_\nu \phi$, for electromagnetic fields $M_{\m\n}=F_{\m\n}$,
and so on. As pointed out above, such a construction does not fit
within the gravitational plus matter splitting of metric theories of
gravity and this suggests that theories of that type should be at
some point in conflict with the experimental evidence supporting the
equivalence principle. Nonetheless, Vollick showed that for
$M_{\m\n}=\partial_\mu\phi \partial_\nu \phi$ the theory field
equations boil down to those of General Relativity (GR) minimally
coupled to a free scalar field. For electromagnetic and fermionic
fields, at leading order in perturbations, one can see that the
usual Einstein-Maxwell and Einstein-Dirac systems are recovered,
{which shows that such theories admit an MTG representation at
that order}. This motivates us to go beyond the perturbative
analysis of \cite{Vollick:2005gc} and try to find a complete
representation of the field equations of those theories. Due to the
technicalities involved, in this paper we will concentrate on the
electromagnetic case, for which an exact representation will be
provided. We would like to point out that the
gravitational-electromagnetic determinantal action has the appealing
property that it combines linearly the first derivatives of the
affine connection with the first derivatives of the $U(1)$
electromagnetic connection. Thus, in some sense, it is treating in a
similar footing spin 1 and spin 2 massless bosons. Understanding the
resulting non-perturbative dynamics of such a theory is thus a
nontrivial question that deserves genuine attention.

In this work we will show that for the gravitational-electromagnetic
case it is possible to find an explicit representation of the full
field equations, and that they turn out to admit a reformulation
that fits within the family of metric theories of gravity, breaking
in this way the apparent conflict posed by the initial
representation of the theory. In fact, the
gravitational-electromagnetic case turns out to be equivalent, at
all orders, to Einstein's gravity coupled to a nonlinear
electrodynamics (NED) theory which, up to a sign, coincides with the
Born-Infeld theory. Our results suggest that there might be
alternative ways of consistently coupling matter and gravity which
break the standard paradigm set by MTGs.

The paper is organized as follows. {In Section \ref{sec:II} we set
the determinantal Born-Infeld-like Lagrangian describing the
gravitational-electromagnetic system, and display the dynamics
resulting in the metric-affine (\`{a} la Palatini) formalism. We show
that the affine connection turns out to be the Levi-Civita
connection. Once the connection has been solved, one can proceed in
two ways. On the one hand one can take advantage of the experience
gained with Born-infeld-like Lagrangians to get the dynamics in a
few steps, as done in Section \ref{BItheory}. However, this
procedure leads to dynamical equations where gravity and
electromagnetism are very intermingled. On the other hand, one can
rework the dynamics to prove that, in spite of the appearances, the
system evolves exactly as expected in an MTG theory, as shown in
Section \ref{MTG}. In Section \ref{SSS} we solve the dynamics for a
charged static black hole to {illustrate aspects of the general
behaviour}. In Section \ref{conclusion} we display the conclusions.}

\section{Model and equations} \label{sec:II}
The theory we will be dealing with can be written in compact form as
\begin{equation}\label{action} I \propto \int dx^4
\left[\sqrt{-q}  - \l \sqrt{-g}\right]
\end{equation}
where $g$ is
the determinant of the space-time metric $g_{\mu\nu}$, and $q$ the
determinant of a tensor $q_{\mu\nu}$, defined as
\begin{equation}\label{eq:defq}
q_{\m\n}\equiv g_{\m\n}+ \e R_{(\m\n)}(\G) +\b F_{\m\n}(A)\ ,
\end{equation}
an object containing the geometric as well as the matter
(electromagnetic) fields. Therefore, besides the metric tensor and
the symmetric part of the Ricci tensor,
$R_{\mu\nu}(\G)\equiv\partial_{\alpha}\Gamma^{\alpha}_{\nu\mu}
-\partial_{\nu}\Gamma^{\alpha}_{\alpha\mu}+\Gamma^{\alpha}_{\alpha\beta}
\Gamma^{\beta}_{\nu \mu}
-\Gamma^{\alpha}_{\nu\beta}\Gamma^{\beta}_{\alpha\mu}$, constructed
upon an arbitrary affine connection $\Gamma^\lambda_{\mu\nu}$, we
have the field strength $F_{\m\n}\equiv \p_{\m} A_{\n}- \p_{\n}
A_{\m}$ of an Abelian ($U(1)$) matter vector field $A_\mu$. The
constant coefficients $\e$ and $\b$ are universal scales with square
length and inverse of $F$ units, respectively. A value of $\l$
different from $1$ implies the cosmological constant
$\Lambda\equiv(1-\lambda)/\e$.

As the connection $\Gamma $ is {\it a priori} assumed to be
independent of the space-time metric (metric-affine formalism), we
will work under the only assumption of the existence of an inverse
for $q_{\m\n}$, {\it i.e.} an object $q^{\m\n}$ defined through the
relation $q^{\m\a}q_{\a\n}\equiv{\d^{\m}}_{\n}$. Performing the full
variation of the action  \eqref{action} one gets
\begin{equation}\label{daction}
\d I \propto \int dx^4
\left[\sqrt{-q} q^{\n\m}\d q_{\m\n}  - \l \sqrt{-g} g^{\m\n}\d
g_{\m\n}\right]=0\ ,
\end{equation}
where $\d q_{\m\n}=\d g_{\m\n} +\e\, \d R_{(\m\n)}(\G)+\b\, \d
F_{\m\n}$, and we call your attention to the index ordering of the
object $q^{\n\m}$ (transpose of $q^{\m\n}$), on which we are not
allowed to make any assumptions about its structure or
symmetries~\footnote{Recall that for any matrix $m$ with inverse
$\bar m$, one has $\d \ln|\det m_{\m\n}|=\bar m^{\n\m}\d
m_{\m\n}$.}. Nonetheless, nothing prevents us from exploiting the
known symmetries of $g_{\m\n}$, $R_{(\m\n)}$ and $F_{\m\n}$, to
rewrite \eqref{daction} as
\begin{equation}\label{daction2}
\int\! dx^4 \left[\(\sqrt{-q} q^{(\m\n)}- \l \sqrt{-g} g^{\m\n} \)\d
g_{\m\n}+\e \sqrt{-q} q^{(\m\n)} \d R_{(\m\n)}(\G) - \b \sqrt{-q}
q^{[\m\n]} \d F_{\m\n} \right] =0\ ,
\end{equation}
where $\d F_{\m\n}= 2\, \p_{[\m} \d A_{\n]}$. {The three
variations are independent, so let us firstly focus on} the
$\G$-related term. The most general form of the variation of the
Ricci tensor is given by
\begin{equation}\label{dRicci}
\d R_{\m\n}= \nabla^{\Gamma}_\a (\d
\G^\a_{\n\m}) -\nabla^{\Gamma}_\n (\d \G^\a_{\a\m}) - S^\th_{\ \n\l}
\d \G^\l_{\th\m} \ ,
\end{equation}
where $S^\th_{\ \a\b}\equiv 2 \G^\th_{[\a\b]}$ is the torsion
tensor. After integrating by parts the $\d R_{\m\n}$ term in
\eqref{daction2} and discarding surface terms \footnote{Regarding
the treatment of surface terms in the metric-affine formalism see
the discussion in \cite{Gomez:2020rnq}.} we obtain the form of the
independent $\delta \Gamma^\a_{\b\m}$ term, which leads to the
equation
\begin{equation}\label{dGamma}
-\nabla^{\Gamma}_\a (\sqrt{-q} q^ {(\m\n)} \d^{\b}_{\ \n}) +
\nabla^{\Gamma}_\n (\sqrt{-q} q^ {(\m\n)} \d^{\b}_{\ \a})-\sqrt{-q}
q^ {(\m\n)} S^\b_{\n\a}=0 \ .
\end{equation}
Tracing this expression we get $\nabla^{\Gamma}_\n (\sqrt{-q} q^
{(\m\n)} ) =\sqrt{-q} q^ {(\m\n)} \tfrac13 S^\a_{\m\a}$. Now, as
shown in \cite{Afonso:2017bxr} -and further discussed in
\cite{BeltranJimenez:2019acz}- theories based on the symmetric part
of the Ricci tensor are projectively invariant, and the torsion
appears only as a projective mode which can be gauged
away.~\footnote{As also pointed out in \cite{Afonso:2017bxr},
``Theories containing the full Ricci tensor will still have a pure
gradient projective symmetry, {\it i.e.}, they are invariant under a
projective transformation. This already suggests that giving up on
the projective symmetry and allowing for the general Ricci tensor
will make the projective mode to become a `Maxwellian field' '' --
see however Ref.~\cite{jimenezdelhom2020}.} This implies that,
without loosing generality, we can choose $\nabla^{\Gamma}_\n
(\sqrt{-q} q^ {(\m\n)} ) =0$, which simplifies the above equation
\eqref{dGamma}.

Gathering all the field equations we have
\begin{eqnarray}
\label{eq:eqs}
\d g_{\m\n}:& \q \sqrt{-q} q^{(\m\n)}- \l \sqrt{-g} g^{\m\n}&=0 \ , \label{dg}\\
\d \G^\a_{\m\n}:&\qq\q\nabla^{\G}_{\a} \(\sqrt{-q} q^{(\m\n)}\)&=0 \ ,\label{dG}\\
\d A_{\m}:&\qq \p_{\m}\(\sqrt{-q} q^{[\m\n]} \)&=0 \ .\label{dA}
\end{eqnarray}
Equation \eqref{dg} implies that on shell, Eq.~\eqref{dG} becomes
$\nabla^{\G}_{\a} \(\sqrt{-g} g^{\m\n}\)=0$, thus fixing the
connection to be Levi-Civita with respect to $g_{\m\n}$
($\nabla^{\G}_{\a}g_{\m\n}=0$), exactly as in GR.

Equations \eqref{dg} and \eqref{dA} govern the dynamics of the
geometry and the electromagnetic field. In principle, they require
inverting the tensor $q_{\m\n}$, and splitting the result in its
symmetric and antisymmetric parts. Noticeably, the relation between
the inverse tensor $q^{\m\n}$ and the other fields can be very
involved, as we will show later. For the moment, and in order to
gain some insight on the role and properties of these equations, we
 find it useful to consider a small excursion and have a glance
at the Born-Infeld (BI) electromagnetic theory first.

\section{Lessons from Born-Infeld nonlinear electrodynamics}\label{BItheory}

Born-Infeld electrodynamics is {dictated} by the Lagrangian
{density} $L_{BI}= -\frac{b^2}{4\pi} \left[\sqrt{-\det
(g_{\m\n}+b^{-1} F_{\m\n})}-\sqrt{-g}\right]$,  the BI constant $b$
having units of electromagnetic field. The dynamical equations read
\begin{equation}\label{BIequation}
\p_{\m}\(\sqrt{-g}\, \FF^{\m\n} \)=0 \ ,
\end{equation}
where  the tensor $\FF$ is given by
\begin{equation}\label{FF}
\FF^{\n\r}\equiv\frac{F^{\n\r}-b^{-2} P \dual
F^{\n\r}}{\sqrt{1+{b^{-2}2S}-{b^{-4}P^2}}}\ .
\end{equation}
Here $S,\, P$ are the scalar and pseudo-scalar field invariants
\begin{equation}\label{SP} S\equiv\tfrac14 F_{\r\l}
F^{\r\l}=-\tfrac14 \dual F_{\r\l} \dual F^{\r\l} \,, \qq  P
\equiv\tfrac14 \dual  F_{\r\l} F^{\r\l} =\tfrac14\, F_{\r\l} \dual
F^{\r\l} \ ,
\end{equation}
 which take part in the relations
\begin{equation}\label{SPrels}
F_{\n\l}  F^{\m\l} - \dual F_{\n\l} \dual F^{\m\l}= 2\, S\,
\d^{\m}_{\n}\ ,  \qq\qq F_{\n\l} \dual F^{\m\l}= \dual F_{\n\l}
F^{\m\l}= P\, \d^{\m}_{\n}\ .
\end{equation}
We note that in terms of these invariants the Lagrangian density can
also be written as
\begin{equation}
L_{BI}= -\frac{b^2}{4\pi}
\sqrt{-g}\left[\sqrt{1+{b^{-2}2S}-{b^{-4}P^2}}-1\right] \
.\label{BI2}
\end{equation}

The stress-energy tensor is {defined, as usual, by} varying
the Lagrangian with respect to the metric,
\begin{equation}\label{TBIdef}
T_{BI}^{\m\n}\equiv \frac{-2}{\sqrt{-g}} \frac{\d L_{BI}}{\d
g_{\m\n}}=\(\frac{- b^{2}}{4\pi}\)
\left[\frac{-2}{\sqrt{-g}}\frac{\d \sqrt{-\det (g_{\m\n}+b^{-1}
F_{\m\n})}}{\d g_{\m\n}}+g^{\m\n} \right]
\end{equation}
(signature $+---$), which results in
--see, for instance, \cite{Ferraro:2010}--
\begin{equation}\label{TBIF}
4\pi b^{-2}\, T_{BI}^{\m\n}=- b^{-2} F^{\m}_{\ \r}
\,\FF^{\n\r} - g^{\m\n}\(1- \sqrt{1+{b^{-2}\,2\,S}-{b^{-4}\, P^2}}\) \ .
\end{equation}

\smallskip
By rearranging the expression \eqref{TBIdef}, the variation of the
squared root determinant of the combined (metric $+$ electromagnetic
field strength) tensors can be expressed as
\begin{equation}\label{dBIdg}
\frac{\d  \sqrt{-\det (g_{\m\n}+b^{-1} F_{\m\n})}}{\d g_{\m\n}}
=\frac{\sqrt{-g}}{2} \(4\pi b^{-2}\,T_{BI}^{\m\n}+g^{\m\n} \)\ .
\end{equation}

\subsection{Dynamical equations for the electromagnetic field}\label{BIdynem}
{The above formulae can be exploited as a direct way to
derive Eqs.~\eqref{dg} and \eqref{dA}. Their usefulness} becomes
apparent when introducing the notation
\begin{equation}
\GG_{\m\n}\equiv g_{\m\n}+ \e R_{(\m\n)} \ ,
\end{equation}
in terms of which we can write $q_{\m\n}= \GG_{\m\n}+ \b F_{\m\n}$.
As a consequence, Eq.~\eqref{dA} is nothing but the BI equation for
the electromagnetic field in a background metric $\GG_{\m\n}$,
namely {(see Appendix \ref{apndxA} for details)}
\begin{equation}\label{VLCBUE}
\p_{\m}\(\sqrt{-\GG}\, \wt\FF^{\m\n} \)=0  \ ,
\end{equation}
where the tilde indicates that the indices are raised by means of
$\wt{\GG}^{\m\n}$, which is the inverse of the ``metric''
${\GG}_{\m\n}$, namely,
\begin{equation}\label{Fstilde}
\wt F^{\a\b}\equiv\wt{\GG}^{\a\m}\wt{\GG}^{\b\n}
F_{\m\n}\,, \qq \dual\wt F^{\a\b}\equiv \tfrac12
\,\tilde\ve^{\,\a\b\r\l} F_{\r\l}= -\tfrac12 \,(-\GG)^{-1/2}
\e^{\a\b\r\l} F_{\r\l} \ .
\end{equation}
($\e^{0123}=1$), from which one obtains the related invariants $\tilde
S,\ \tilde P$, and the corresponding $\wt\FF^{\m\n}$ tensor
\begin{equation}\label{FFtilde}
\wt\FF^{\n\r}\equiv\frac{\wt
F^{\n\r}-\b^{2} \tilde P \dual\wt F^{\n\r}}{\sqrt{1+{\b^{2}2 \tilde
S}-{\b^{4}\tilde P^2}}}\ .
\end{equation}


\subsection{Dynamical equations for the geometry}\label{BIdyngeom}
Equation \eqref{dg}, the variation of action~\eqref{action} with
respect to the metric $g_{\m\n}$, can also be investigated by
exploiting the results of the Born-Infeld theory. As
$\frac{\d\GG_{\a\b}}{\d g_{\m\n}}=\d^{\m}_{\a}\d^{\n}_{\b}\,$, we
can write
\begin{equation}\label{daction4}
\frac{\d \sqrt{-q}}{\d g_{\m\n}}=\frac{\d \sqrt{-q}}{\d
\GG_{\m\n}}=\frac{\d \sqrt{-\det (\GG_{\m\n}+ \b F_{\m\n})}}{\d
\GG_{\m\n}} \ ,
\end{equation}
and using \eqref{dBIdg}, the dynamical equation \eqref{dg} reads
\begin{equation} \label{daction5}
\sqrt{-\GG} \(\wt{\GG}^{\n\m}+
4\pi \b^{2}\,\wt T_{BI}^{\m\n}\)=\l\sqrt{-g}\, g^{\m\n}\ ,
\end{equation}
where $\wt T_{BI}^{\m\n}$ is the tilded version of
the stress-energy tensor \eqref{TBIF}. This means that $g^{\m\n}$
must be replaced with $\wt{\GG}^{\m\n}$, and the tilded magnitudes
mentioned in \eqref{Fstilde} must enter into play. Besides, $b^{-1}$
is replaced by $\b$.

Contracting this expression with $\GG_{\l\n}$, and substituting the
determinant $\sqrt{-\GG}=\sqrt{-g}\sqrt{\det(\d^{\m}_{\ \n} + \e
R^{\m}_{\ \n})}$,  where $R^{\m}_{\ \n}=g^{\m\r} R_{\r\n}$ {
is written with the Levi-Civita connection}, the dynamical equations
for the geometry become
\begin{equation}\label{Einstein}
\l\, \frac{\d^{\m}_{\ \n} + \e R^{\m}_{\ \n}}{\sqrt{\det(\d^{\m}_{\
\n} + \e R^{\m}_{\ \n})}} = \d^{\m}_{\ \n}+4\pi \b^{2}\,\wt
T^{\m}_{\ \n\, \, BI} \ ,
\end{equation}
where
\begin{equation}\label{TBIFud}
4\pi \b^{2}\, \wt{T}^{\m}_{\
\n\, \, BI}=- \b^{2}\, \wt{\FF}^{\m\r}\, \wt F_{\n\r}  -
\d_{\,\n}^{\m}\(1- \sqrt{1+{\b^{2}\,2\,\tilde S}-{\b^{4}\, \tilde
P^2}}\)\ .
\end{equation}

Notice that de Sitter geometry, or any other geometry such that its Ricci
tensor is $R_{\,\n}^{\m}=-\Lambda\,\d_{\,\n}^{\m}$ ($\Lambda$ is the
cosmological constant), is a vacuum solution to Eq.~\eqref{Einstein}
provided that $\l$ is chosen to be
\begin{equation}\label{Lambda}
\l=1-\e\Lambda \ .
\end{equation}
Note that Einstein's equations are recovered from
Eq.~\eqref{Einstein} in the weak field regime. This is a foreseeable
behavior since the action \eqref{action} becomes the usual
Einstein-Maxwell action in such a limit. As $\sqrt{\det (\delta
_{\,\n }^{\m }+\e R_{\,\n }^{\m} )} \simeq 1+\frac{\e }{2}R$,
Eq.~\eqref{Einstein} goes to
\begin{equation} \l\ \d^{\m}_{\
\n}+\l\e \left(R^{\m}_{\ \n}-\frac{R}{2}~\d^{\m}_{\ \n}\right)
\simeq \d^{\m}_{\ \n}+4\pi \b^{2}\,T_{\ \n\ Maxwell}^{\m} \ ,
\end{equation}
{\it i.e.},
\begin{equation}\label{approx}
R^{\m}_{\ \n}-\frac{R}{2}~\d^{\m}_{\ \n}-\Lambda ~\d^{\m}_{\
\n}\simeq 4\pi\b^2\e^{-1}\, T_{\ \n\ Maxwell}^{\m} \ ,
\end{equation}
where $\l R^{\m}_{\,\n}=(1-\e\Lambda) R^{\m}_{\,\n}$ has been
approximated by $R^{\m}_{\,\n}$ because both, the curvature and the
cosmological constant must be weak in Eq.~\eqref{approx}. Besides,
this equation shows that the Newton constant emerges from the
relations between the universal scales in the action as
\begin{equation}\label{k2}
\k^2\equiv 8\pi\, G=4\pi\b^2\e^{-1}
\end{equation}

\section{MTG behaviour of the dynamical equations}\label{MTG}
Equations \eqref{VLCBUE} and \eqref{Einstein}, which have been
obtained by a straightforward variation of the action, seem to imply
that the equivalence principle is violated by the action
\eqref{action}. This conclusion comes from the presence of the Ricci
tensor $R_{(\m\n)}$ in the Eq.~\eqref{VLCBUE} governing the dynamics
of the electromagnetic field; in fact, the Ricci tensor enters in
the volume $\sqrt{-\GG}$ and in the tensor ${\wt
F}^{\m\n}=\GG^{\m\l}\GG^{\n\r}F_{\l\r}$. Moreover, the source of
curvature in the r.h.s. of Eq.~\eqref{Einstein} is also contaminated
with the Ricci tensor. However, we will show that the dynamical
equations can be rearranged in such a way that both contaminant
effects of the Ricci tensor will disappear,
{eliminating the need to resort to the ``tilde operation'' or to the volume $\sqrt{-\GG}$. }
This means that the dynamics will reveal its MTG character, despite the fact that
the action does not explicitly exhibit such feature.

\subsection{Electrodynamics}
Let us come back to the Eq.~\eqref{dA}, where $q^{[\m\n]}$ is the
antisymmetric part of the tensor inverse of $q_{\m\n}=g_{\m\n}+\e
R_{(\m\n)}+\b F_{\m\n}$.\footnote{$q^{\m\n}$ is not meant to
represent $q_{\m\n}$ with its indices raised with $g^{\mu\nu}$, {\it
i.e.}, $q^{\m\n}\neq g^{\mu\alpha}g^{\nu\beta}q_{\alpha\beta}$.} In
order to obtain $q^{[\m\n]}$, let us firstly use Eq.~\eqref{dg} to
write
\begin{equation}\label{qup}
q^{\m\n}=q^{(\m\n)}+q^{[\m\n]}  =\g\,g^{\m\a} \(
\d_{\a}^{\ \n} + a_{\a}^{\ \n} \) \ ,
\end{equation}
where we have introduced the notation $\g \equiv \l\sqrt{{g}/{q}}\,$
and $a_{\a}^{\ \,\n}\equiv \g^{-1}g_{\a\b} q^{[\b\n]}\,$, such that
$a^{\mu\nu}= g^{\mu\alpha}a_{\alpha}^{\ \nu}=\g^{-1}q^{[\mu\nu]}$
and $a_{\mu\nu}=
\g^{-1}g_{\mu\alpha}g_{\nu\beta}q^{[\alpha\beta]}=-a_{\nu\mu}$. Now
let us introduce the matrix $\hat\Omega$ defined as
\begin{equation}\label{00}
q_{\mu \nu }=\gamma ^{-1}~g_{\mu \lambda }~\Omega _{~\nu }^{\lambda
} \ .
\end{equation}
For $q_{\m\n}$ to be the inverse of $q^{\m\n}$ it must be
\begin{equation}
\delta _{\nu }^{\mu }=q^{\mu \alpha }q_{\alpha \nu
}=\Omega _{~\nu }^{\mu }+a_{~\lambda }^{\mu }~\Omega _{~\nu
}^{\lambda }\ ,
\end{equation}
which can be rewritten as a matrix equation for $\hat{\Omega}$,
\begin{equation}\label{03}
\hat{I}~=\hat{\Omega}+\hat{a}\cdot
\hat{\Omega}\ .
\end{equation}
From here the matrix $\hat\Omega$ can be solved in terms of the
antisymmetric matrix $\hat a$, thus linking $q_{\m\n}$ in
Eq.~\eqref{00} to $q^{[\m\n]}$ in Eq.~\eqref{qup}. According to
Eq.~\eqref{03}, $\hat\Omega$ might result in a matrix written in
terms of $\hat a$ and its dual $\dual \hat a$
{($\dual a_{\a\b}=\tfrac12\, \varepsilon_{\a\b\r\l}~ a^{\r\l}=\tfrac12\,
(-g)^{1/2}\epsilon_{\a\b\r\l}~a^{\r\l}$ ).}
Paying attention to the relations
\begin{equation}\label{04}
-\hat{a}\cdot \hat{a}+\dual\hat{a}\cdot \dual\hat{a}=2~s~\hat{I},~~\
\ \ \ \ \ ~~\ \ ~\ -\hat{a}\cdot \dual\hat{a}=-\dual\hat{a}\cdot
\hat{a} = p~\hat{I}~~,
\end{equation}
where $s$, $p$ are the scalar and pseudo-scalar associated with
$\hat a$,
\begin{equation}
s \equiv \tfrac14 a_{\a\b}~a^{\a\b}=
-\tfrac14 \dual a_{\a\b} \dual a^{\a\b} \,, \qq p\equiv \tfrac14
a_{\a\b}~\dual a^{\a\b}=\tfrac14 \dual a_{\a\b}~a^{\a\b}\ ,
\end{equation}
one concludes that $\hat{\Omega}$ should have the
form \footnote{A possible term $\hat{a}\cdot \hat{a}$ would be
absorbed in the other ones due to the first of the relations
\eqref{04}.}
\begin{equation}
\hat{\Omega}=c_{1}~\hat{a}+c_{2}~\dual\hat{a}+c_{3}~
\dual\hat{a}\cdot \dual\hat{a}+c_{4}~\hat{I} \ .
\end{equation}
Substituting this expression in Eq.~\eqref{03}, one finds that
\begin{equation}
c_2=-p~c_1~,\ \ \ c_3=-c_1=c_4~,\ \ \ c_1=-\frac{1}{1+2~s-p}\
\end{equation}
(we used $\dual\dual=-1$). In sum, it is
\begin{equation}\label{07}
\hat{\Omega}=\frac{\hat{I}~-\hat{a}+p~\dual\hat{a}+\dual\hat{a}\cdot
\dual\hat{a}}{1+2~s-p^{2}}~.
\end{equation}

Since the antisymmetric part of $q_{\mu\nu}=\gamma ^{-1}g_{\mu
\lambda }~\Omega _{~\nu }^{\lambda }$ is $\beta F_{\mu \nu }$, we
get the following relation between the unknown $\hat a$ and $\hat
F$:
\begin{equation}\label{05}
q_{[\mu \nu ]}=\gamma ^{-1}\frac{-a_{\mu \nu }+p~^{\ast }a_{\mu \nu
}}{1+2~s-p^{2}}=\beta ~F_{\mu \nu }\ .
\end{equation}
So, in order to express $q^{[\m\n]}=\g~a^{[\m\n]}$ in
Eq.~\eqref{dA}, we must solve the former equation for $a^{[\m\n]}$.

Let us have a short break here to say a few words about
$\g\equiv\l\sqrt{g/q}$. The determinant of Eq.~\eqref{00} yields
$q^{-1}=\gamma ^{4}g^{-1}\det (\hat{I}+\hat{a})$; thus it is
\begin{equation}
(\gamma \lambda )^{-2}=\det (\hat{I}+\hat{a}) \
\end{equation}
(use Eq.~\eqref{03}). The experience with Born-Infeld Lagrangians
tells us that $\det (\hat{I}+\hat{a})=1+2~s-p^2$ (since $a_{\m\n}$
is antisymmetric; {\it cf.} Eq.~\eqref{BI2}). Therefore
\begin{equation}\label{gl}
(\g\l)^{-2} = 1+2~s-p^{2} \ .
\end{equation}
Thus, the equation \eqref{05} reduces to
\begin{equation}\label{06}
\g\l~(-a_{\mu \nu }+p~^{\ast }a_{\mu \nu })={\bar F}_{\mu \nu } \ ,
\end{equation}
where ${\bar F}_{\mu \nu }\equiv \b\l^{-1} F_{\mu \nu
}$. By applying the dual operator to this equation, one gets a
second equation to solve $\hat a$ and $\dual\hat a$. The result is
\begin{equation}\label{a}
a_{\m\n}=-(\g\l)^{-1}~\frac{{\bar F}_{\m\n}+p~\dual {\bar
F}_{\m\n}}{1+p^2}\ .
\end{equation}
From Eq.~\eqref{06} one also obtains
\begin{equation}
{\bar S}=\frac{1}{4}{\bar F}_{\rho \lambda }{\bar F}^{\rho \lambda
}=(\g\l)^{2}~(s-2p^{2}-sp^{2})\ ,
\end{equation}
\begin{equation}\label{p}
{\bar P} = \frac{1}{4}{\bar F}_{\rho \lambda }~\dual{\bar F}^{\rho
\lambda } = (\g\l)^{2}~p~(1+2s-p^{2}) = p\ .
\end{equation}
Besides,
\begin{equation}\label{12SP}
1-2{\bar S}-{\bar P}^{2} =1-2(\g\l)^{2}(s-2p^{2}-sp^{2})-p^{2}
=\left[\g\l (1+p^2)\right]^{2}\ .
\end{equation}
Finally, with the help of Eq.~\eqref{dg}, we have obtained the
results that allow us to express Eq.~\eqref{dA} --the field equation
arising from the variation with respect to the gauge field $A_\mu$--
explicitly in terms of the electromagnetic field strengths
$F^{\m\n}$ and $\dual F^{\m\n}$, without the contaminant presence of
the Ricci tensor. In fact $\sqrt{-q}q^{[\m\n]}$ is
\begin{equation}
\sqrt{-q} q^{[\m\n]}=\g^{-1}\l\sqrt{-g}~\g
a^{\m\n}=-\g^{-1}\sqrt{-g}~\frac{{\bar F}^{\m\n}+p~\dual{\bar
F}^{\m\n}}{1+p^2}=-\l\sqrt{-g}~\frac{{\bar F}^{\m\n}+{\bar
P}~\dual{\bar F}^{\m\n}}{\sqrt{1-2{\bar S}-{\bar P}^{2}}}
\end{equation}
Thus Eq.~\eqref{dA} is
\begin{equation}\label{dAF}
\p_{\n} \left[\sqrt{-g}~\frac{\bar F^{\m \n}+ \bar P \dual\bar
F^{\m\n}}{\sqrt{1-2 \bar S- \bar P^2}}\right]=0 \ .
\end{equation}
This dynamical equation resembles the equation \eqref{BIequation} of
the standard Born-Infeld electromagnetic theory. However, $\bar S$
enters the square root with the opposite sign; there is also a
change of sign in the numerator. This negative sign in the square
root of Eq.~\eqref{dAF} has a strong impact on the features of the
solutions because, as noted by Vollick in \cite{Vollick:2005gc}:
`the square root does not, by itself, constrain the magnitude of the
electric field'. This effect will be evidenced in the static
spherically symmetric solution presented in Section \ref{SSS}.

\subsection{Einstein equations}

The result \eqref{07} is also useful to show that the dynamics of
the geometry is dictated by a source free of the presence of the
Ricci tensor. In fact, $q_{(\m\n)}=g_{\m\n}+\e R_{(\m\n)}$ can be
compared with the expression resulting from Eqs.~\eqref{00} and
\eqref{07}. Thus one obtains
\begin{equation}\label{08}
g_{\m\n}+\e R_{(\m\n)}=\g\l^2 \left(g_{\m\n}+\dual a_{\m\a}\dual
a^\a_{\ \n}\right)\ .
\end{equation}
For a vanishing matter field ({\it i.e.}, $a_{\m\a}=0$ and $\g\l=1$)
one gets $\e R_{(\m\n)}=-(1-\l)~g_{\m\n}=-\e\Lambda~ g_{\m\n}$; in
particular, the de Sitter geometry is a vacuum solution. Tracing
Eq.~\eqref{08} it yields
\begin{equation}
4+\e R=\g\l^2 \left(4+4 s\right)\ .
\end{equation}
Therefore, the Einstein tensor $G_{\m}^{\ \n}=R_{\m}^{\ \n}-(1/2) R~
\d_{\m}^{\ \n}$ fulfills the equation
\begin{equation}\label{GEinstein}
\e G_{\m}^{\ \n}=\d_{\m}^{\n}+\g\l^2~\left[\dual a_{\m\a}\dual
a^{\a\n}-(1+2s)~\d_{\m}^{\ \n}\right]\ ,
\end{equation}

Equations \eqref{a}, \eqref{SP} and \eqref{SPrels} imply that
\begin{equation}
\dual a_{\m\a}\dual a^{\a\n}=\frac{2~({\bar S}+{\bar
P}^2)~\d^\n_\m-(1+{\bar P}^2)~{\bar F}_{\m\a}{\bar
F}^{\n\a}}{1-2{\bar S}-{\bar P}^2}\ ,
\end{equation}
therefore
\begin{equation}
1+2s=1+\frac{1}{2} \dual a_{\m\a}\dual a^{\a\m}=\frac{1+3{\bar
P}^2-2{\bar S}{\bar P}^2}{1-2{\bar S}-{\bar P}^2}\ .
\end{equation}
Thus, using Eqs.~\eqref{p} and \eqref{12SP}, Einstein's equations
\eqref{GEinstein} reads
\begin{equation}\label{eG}
\e G_{\m}^{\ \n}=\d_{\m}^{\ \n}-\l~\frac{{\bar F}_{\m\a}{\bar
F}^{\n\a}+(1-2{\bar S})~\d_{\m}^{\ \n}}{\sqrt{1-2{\bar S}-{\bar
P}^2}} \ .
\end{equation}
 As expected, the curvature is sourced just by the matter fields.

\subsection{Equivalent MTG Lagrangian}
At this stage one wonders if there exists an MTG Lagrangian leading
to Eqs.~\eqref{dAF} and \eqref{eG}. So, we will look for a nonlinear
electrodynamics (NED) whose Lagrangian leads to the dynamics
\eqref{dAF}, and whose metric energy-momentum tensor coincides with
the source in Eq.~\eqref{GEinstein}. A generic NED theory has an
action of the form
\begin{equation}
I_{NED}=-\frac{1}{4\pi}\int d^4x \sqrt{-g} \,\varphi(\bar S,\bar P)
\end{equation}
where $\varphi(\bar S, \bar P)$ is an arbitrary function of the
scalar and pseudo-scalar field invariants. The associated
stress-energy tensor is
\begin{equation}\label{TNED}
T^{NED}_{\a\b}=\frac{-2}{\sqrt{-g}} \frac{\d L_{NED}}{\d
g^{\a\b}}=\frac{1}{4\pi}~\left[\varphi_{\bar S}\, {\bar F}_{\a}^{\
\r}{\bar F}_{\b\r} - (\varphi-{\bar P} \varphi_{\bar
P})\,g_{\a\b}\right]\ .
\end{equation}
The differences of sign pointed out in Eq.~\eqref{dAF}, if compared
with Eqs.~\eqref{BIequation} and \eqref{FF}, suggest a
Born-Infeld-like function $\varphi(\bar S, \bar P) = c_{1}\sqrt{1-2
\bar S-\bar P^2}+c_{0}$, with $c_{1}$ and $c_{0}$ constants. Thus,
the stress-energy tensor will result
\begin{equation}
T_{\a\b}=\frac{1}{4\pi} \left[-\frac{c_{1}\ \bar F_{\a}^{\ \r} \bar
F_{\b\r}}{\sqrt{1-2\bar S-\bar P^2}} -\left( c_{1}\sqrt{1-2 \bar
S-\bar P^2}+c_{0}+c_{1}~\frac{{\bar P}^2}{\sqrt{1-2\bar S-\bar
P^2}}\right) ~g_{\a\b}\right] \ .
\end{equation}
Since $4\pi\b^2=\e\k^2$ (see Eq.~\eqref{k2}), then Eq.~\eqref{eG}
can be read as the Einstein's equation $\e G_{\m}^{\ \n}=\e \k^2
T_{\m}^{\ \n}$ provided that the constants are fixed as
\begin{equation}\label{const}
c_{0} = -\b^{-2} \, ,\qq c_{1}=\frac{4\pi\l}{\e\,\k^2}=\l\b^{-2}\ .
\end{equation}
It is thus evident that the field equations of the theory
(\ref{action}) can be also derived from an action that adds a
BI-like action to the Einstein-Hilbert one,\footnote{No matter
whether the variational process is metric-affine or just metric.}

\begin{equation}
I_{MTG}[g_{\m\n},A_\m]=\frac{1}{2\k^2}\int d^4x\sqrt{-g}~R[g_{\m\n}]
-\frac{1}{4\pi\b^2}\int d^4x\sqrt{-g}\left(\l \sqrt{1-2\bar S-\bar
P^2} - 1\right)\ .
\end{equation}
In the weak field limit, this action goes to
\begin{equation}
I_{MTG}\rightarrow \frac{1}{2\k^2}\int d^4x\sqrt{-g}
\left(R+2\Lambda\right) +\frac{1}{16\pi\l}\int d^4x\sqrt{-g}~
F_{\r\l}F^{\r\l}\ ,
\end{equation}
with $\Lambda=(1-\l)\e^{-1}$.

\section{Spherically symmetric solution}\label{SSS}

Born-Infeld electrostatics is well known for avoiding the divergence
of the field of a point charge in a flat background. One could
reasonably expect that the action \eqref{action} will soften both
the geometric and the electric singularities. However, as already
said, any geometry such that
$R_{\,\n}^{\m}=-\Lambda\,\d_{\,\n}^{\m}$ is a solution to the
equations \eqref{Einstein} in the absence of sources. In particular,
Schwarzschild-de Sitter geometry is the spherically symmetric vacuum
solution. So, contrary to what could be expected, the action
\eqref{action} does not remove the geometric singularity. Let us
then turn {to the equations \eqref{VLCBUE} and
\eqref{Einstein} (or, equivalently, \eqref{dAF} and \eqref{eG}) to
know the consequences of the interaction between gravity and
electromagnetism in the present theory.}

\smallskip
We start by proposing a spherically symmetric configuration where
the electrostatic field is characterized by an unknown function
$e(r)$ which depends on the radial coordinate,
\begin{equation}
F=\tfrac{1}{2}\, F_{\mu \nu }~dx^{\mu }\wedge dx^{\nu
}=e(r)~dt\wedge dr=-\sqrt{\b Q}~u^{-2}~e(u)~dt\wedge du~,
\end{equation}
($u\equiv \sqrt{\b Q}~r^{-1}$ is non-dimensional and the electric
charge $Q$ is assumed to be positive), and the interval has the form
\begin{equation}
ds^{2}=f(r)~dt^{2}-\frac{dr^{2}}{f(r)}-r^{2}(d\theta ^{2}+\sin
^{2}\theta ~d\phi ^{2})=f(u)~dt^{2}-\frac{\b
 Q}{u^{4}}~\frac{du^{2}}{f(u)}-\frac{\b Q}{u^{2}}~(d\theta ^{2}+\sin
^{2}\theta ~d\phi ^{2})~.\label{metric}
\end{equation}
The function $f(u)$ can be conveniently written as
\begin{equation}\label{g00}
g_{00}=f(u)=1+\frac{\b Q}{\e }\,u\,h(u)\ .
\end{equation}
Thus, the tensor $\delta _{~\n }^{\m}+\e R_{~\n }^{\m }$ in the
chart $(t,u,\theta ,\phi )$ turns out to be
\begin{equation}
\delta _{~\n }^{\m}+\e R_{~\n }^{\m }=
\begin{pmatrix}
1+u^{4}~[h^{\prime }(u)+\frac{u}{2}~h^{\prime \prime }(u)] & 0 & 0 & 0 \\
0 & 1+u^{4}~[h^{\prime }(u)+\frac{u}{2}~h^{\prime \prime }(u)] & 0 & 0 \\
0 & 0 & 1-u^{4}~h^{\prime }(u) & 0 \\
0 & 0 & 0 & 1-u^{4}~h^{\prime }(u)
\end{pmatrix}
~,\label{matrix}
\end{equation}
and the square root of the determinant of $\GG_{\mu \nu }=g_{\mu \nu
}+\e R_{\mu \nu }$ reads
\begin{equation}
\sqrt{-\mathcal{G}}=(\b Q)^{3/2}\ \sin \theta\ u^{-4}~\left(
1+u^{4}~[h^{\prime }(u)+\frac{u}{2}~h^{\prime \prime }(u)]\right)
\left( 1-u^{4}~h^{\prime }(u)\right)\ .  \label{det}
\end{equation}

Equation \eqref{VLCBUE} is fulfilled for any function $h(u)$
whenever the field $e(u)$ is
\begin{equation}
e(u)=\b^{-1}~u^{2}~\frac{1+u^{4}~[h^{\prime}(u)+\frac{u}{2}~h^{\prime
\prime}(u)]}{\sqrt{1+u^{4}-u^{4}~h^{\prime }(u)~\left( 2-
{u^{4}}~h^{\prime }(u)\right) }} \ . \label{efield}
\end{equation}
In fact, in such a case the expression
$\sqrt{-\GG}\,\wt{\mathcal{F}}^{\nu \varrho }$ in Eq.~\eqref{VLCBUE}
does not depend on $u$ besides a global sign change at the value of
$u$ where $u^{4}~h^{\prime }(u)=1$; if $u^{4}~h^{\prime }(u)<1$,
then
\begin{equation}
\sqrt{-\GG}\ \wt{\mathcal{F}}=Q\,\sin\theta\ \frac{\p}{\p t}\wedge
\frac{\p}{\p u}\ .
\end{equation}

Remarkably, Eq.~\eqref{efield} shows that the well known regular
Born-Infeld point-like solution $e(u)=\b^{-1} u^{2}(1+u^{4})^{-1/2}$
is a valid solution not only for a Minkowskian background, but for a
fixed background space-time where $h(u)$ is a constant
(Schwarzschild geometry). Moreover,
\begin{equation}
e(u)=\frac{\b^{-1}~u^{2}~\l}{\sqrt{\l^2+u^{4}}}~.
\label{pointcharge}
\end{equation}
is the solution in the Schwarzschild-de Sitter background, which
corresponds to
\begin{equation}
h(u)=\frac{\l-1}{3\, u^{3}}+ \text{constant}=-\frac{\e \Lambda }{3\,
u^{3}}+ \text{constant}\ .
\end{equation}

Let us now focus on Eq.~\eqref{Einstein}. The stress-energy tensor
for the proposed configuration takes the form
\begin{equation}
\label{tud} \wt T^{\m}_{\ \n\ BI}=-\frac{1}{4\pi\b^2}\
\mathrm{diag}(1-C(u),\ 1-C(u),\ 1-C(u)^{-1},\ 1-C(u)^{-1})\ ,
\end{equation}
where
\begin{equation}
C(u)=\frac{\sqrt{1+u^{4}-u^{4}~h^{\prime }(u)\ (2-u^{4}~h^{\prime
}(u))}}{1-u^{4}~h^{\prime }(u)}\ .\label{g}
\end{equation}

\noindent Function $h(u)$ appears only through its first and second
derivatives in Eqs.~\eqref{matrix} and \eqref{g}. The solution to
``Einstein equations'' \eqref{Einstein} is
\begin{equation}
h(u)=\int ~du~u^{-4}\, \left( 1-\sqrt{\l^{2}-u^{4}}\right) \ .
\label{h}
\end{equation}
At this point we must remember that the field
$\sqrt{-\GG}\,\wt{\mathcal{F}}^{\m \n }$ in Eq.~\eqref{VLCBUE} is
well defined if $u^{4}~h^{\prime }(u)<1$. In the light of the result
\eqref{h}, this implies that both the metric and the electromagnetic
field are well defined whenever $u$ belongs to the interval
$0\leqslant u<\sqrt{|\l|}$.

By replacing the result \eqref{h} into Eq.~\eqref{efield}, $e(u)$
turns out to be
\begin{equation}\label{09}
e(u)=\frac{\b^{-1}~u^{2}~\l}{\sqrt{\l^{2}-u^{4}}} ~.
\end{equation}
Therefore, if the geometry is not a mere background but is sourced
by the electrostatic field, then $e(u)$ {must lose its
smoothness in order to be able to solve the full set of equations;
$e(u)$ is} singular at $u=\sqrt{|\l|}$. The different signs in the
square roots of Eqs.~\eqref{pointcharge} and \eqref{09} is a direct
consequence of the negative sign accompanying $\bar S$ in
Eq.~\eqref{dAF}, contrasting with the positive sign in BI theory.
The singularity at $u=\sqrt{|\l|}$ is shared by both the geometry
and the electrostatic field, as evidenced by the field scalar
invariant and the scalar curvature which yield \footnote{However
$\tilde{S}(u)=-\frac{\b^{-2}~u^{4}}{2~\l^{2}}$ does not diverge.}
\begin{equation}
S=-\frac{e(u)^{2}}{2}=\frac{\b^{-2}~u^{4}~\l^{2}}{\l^{2}-u^{4}}~,~\
\ \ ~\ \ \ \ ~~~\ \ \ R=R_{\mu }^{\mu }=4~\e ^{-1}\left(
\frac{\l^{2}-\frac{u^{4}}{2}}{\sqrt{\l^{2}-u^{4}}}-1\right) ~.
\end{equation}

The geometric character of the singularity is not only evidenced by
the divergent values of $S$ and $R$, but in the finite proper time
required for a particle to reach $u=|\l|^{1/2}$. In fact, the metric
is well behaved at the singularity, since the function $h(u)$ in
Eq.~\eqref{h} is regular at $u=|\l|^{1/2}$.

Let us analyze the function $h(u)$, which is the basic block of
$g_{00}$ as shown in Eq.~\eqref{g00}. By expanding the integrand in
Eq.~\eqref{h}, one gets the behavior of the function $h(u)$ at the
lowest orders; using Eq.~\eqref{Lambda}, the result is
\begin{equation}
h(u)=-\sqrt{\b}\, Q^{-3/2}M-\frac{\e \Lambda }{3\,
u^{3}}+\frac{u}{2\,\l}+O(u^{5})~,
\end{equation}
where $M$ is the integration constant representing the mass.
According to Eq.~\eqref{g00}, $g_{00}$ is
\begin{equation}\label{g00aprox}
g_{00}=f\left(u=\frac{\sqrt{\b Q}}{r}\right)=1-\frac{\b^2 M}{\e\, r}
-\frac{\Lambda }{3}~r^{2}+\frac{(\b Q)^{2}}{2\, \e \,\l\,  r^2}\,
+\frac{(\b Q)^{4}}{40\, \e \,\l^3\,  r^6}+O(r^{-9})~.
\end{equation}
Thus, Reissner-Nordstrom-de Sitter geometry is recovered (remember
that $\b^{2}\e^{-1} =2 G$); however the term of the electric charge
is altered by the unexpected presence of the cosmological constant.

For $\l=1$ (absence of a cosmological constant) the function $h(u)$
is
\begin{equation}
h(u)=-\sqrt{\b}\, Q^{-3/2}M-\frac{1}{3~u^{3}}\left[1-3\,
\sqrt{1-u^4}+2 \
{_{2}}F_{1}\(-\tfrac{3}{4},\tfrac{1}{2};\tfrac{1}{4};u^4\)\right] \,
,
\end{equation}
where $_{2}F_{1}$ is the hypergeometric function. According to
Eq.~\eqref{g00}, to search for horizons we have to solve
\begin{equation}\label{hor}
0=1+\frac{\b Q}{\e }~u_H~h(u_H)\quad\Rightarrow\quad \frac{\e}{\b\,
Q}=\sqrt{\b}\, Q^{-3/2}M\, u_H+\frac{1}{3~u_H^{2}}\left[1-3\,
\sqrt{1-u_H^4}+2 \
{_{2}}F_{1}\(-\tfrac{3}{4},\tfrac{1}{2};\tfrac{1}{4};u_H^4\)\right]\
.
\end{equation}
The function in the r.h.s. of the Eq.~\eqref{hor} is plotted in
Figure 1 for different values of the non-dimensional parameter
$\alpha\equiv\sqrt{\b}\, Q^{-3/2}M$. The horizons are found by
equaling the function with the positive constant $\frac{\e}{\b\, Q}$
in the interval $0<u<1$. We thus recognize that, depending on the
relation between both parameters, the geometry can have two, one, or
no horizons.

\begin{figure}
\includegraphics[width=.5\textwidth]{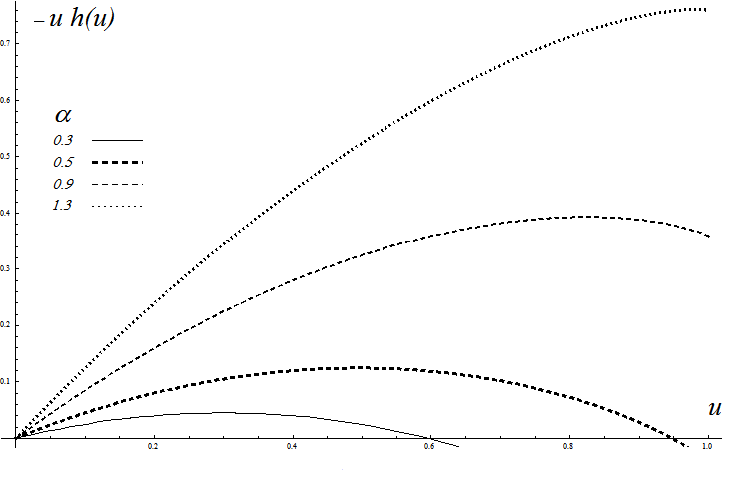}
\caption{The function $-u\, h(u)=\alpha\,
u+\frac{1}{3~u^{2}}\left[1-3\, \sqrt{1-u^4}+2 \
{_{2}}F_{1}\(-\frac{3}{4},\frac{1}{2};\frac{1}{4};u^4\)\right]$ for
different values of $\alpha$.}
\end{figure}

\bigskip
It is useful to compare our results with the solution to the usual
Einstein-Born-Infeld (EBI) dynamics \cite{Plebanski:1984}, which we
summarize below. The functions $e(u)$ and $h(u)$ are
\begin{equation}\label{EBI}
e_{EBI}=\frac{\b^{-1}~u^{2}~\l}{\sqrt{1+\l^{2} u^{4}}} ~, \quad\quad
h_{EBI}=-\int ~du~u^{-4} \left( \l-\sqrt{1+\l^{2} u^{4}}\right) \ .
\end{equation}
The geometry is singular at $u=\infty$ ($r=0$), and there can be
two, one or no horizons depending on the relations between the
parameters (see details in Ref.~\cite{Breton:2002}). However the
field invariant $S$ remains bounded,
\begin{equation}
S_{EBI}=-\frac{\b^{-2}~u^{4}~\l^2}{2(1+\l^{2} u^{4})}~.
\end{equation}
In the weak field region the metric takes the form
\begin{equation}
g_{00\ EBI}=1-\frac{\b^2 M}{\e\, r} -\frac{\Lambda
}{3}~r^{2}+\frac{(\b Q\l)^{2}}{2 \e \,  r^2}-\frac{(\b Q\l)^{4}}{40
\e \, r^6}+O(r^{-9})~,
\end{equation}
which exhibits subtle differences with respect to
Eq.~\eqref{g00aprox} coming from different signs and the different
role of $\l$ in $h_{EBI}(u)$ as compared to $h(u)$ (see
Eqs.~\eqref{h} and \eqref{EBI}).

\section{Summary and Conclusion}\label{conclusion}

{We have studied the field equations coming from a Born-Infeld-type
determinantal Lagrangian that linearly combines gravity and matter,
when varied within a metric-affine (\`{a} la Palatini)} framework. This
formulation explicitly violates the postulates of metric theories of
gravity (MTG) by construction, which raised concerns about its
compatibility with the Einstein equivalence principle. However, by
carefully analyzing the field equations when the matter sector is
described by an electromagnetic field, we have shown that the
dynamics of the theory can be equivalently obtained from an MTG
action which adds the (Einstein-Hilbert) GR action to a Born-Infeld
electrodynamics theory with a ``{\it wrong}'' sign. This result is
exact and independent of the symmetries of the particular solution
presented in Section \ref{SSS}.  The appearance of a non-standard
sign in the electromagnetic sector prevents the bound of the
corresponding invariants, which leads to undesired physical
properties.

It is important to note that the methods introduced here to deal
with the inverse of the tensor $q_{\mu\nu}$ are generic and can be
applied whenever an antisymmetric part appears in the linear
combination that defines $q_{\mu\nu}$. Thus, more general
electromagnetic scenarios and combinations of different matter
fields can be tackled following a similar procedure. Despite the
fact that the model considered here does not prevent curvature
divergences, the possibility of more general actions that result in
regularized geometric and matter sectors cannot be ruled out. Thanks
to our results, such theories are now accessible to exploration and
new results will be reported elsewhere soon.

\acknowledgments

This work was partially supported by Conselho Nacional de
Desenvolvimento Cient\'{\i}fico e Tecnol\'{o}gico (CNPq), Coordena\c{c}\~{a}o de
Aperfei\c{c}oamento de Pessoal de N\'{\i}vel Superior (CAPES), Consejo
Nacional de Investigaciones Cient\'{\i}ficas y T\'{e}cnicas (CONICET),
Universidad de Buenos Aires, by the Spanish Grant
FIS2017-84440-C2-1-P funded by MCIN/AEI/ 10.13039/501100011033
``ERDF Away of making Europe'', Grant PID2020-116567GB-C21 funded by
MCIN/AEI/10.13039/501100011033, the project PROMETEO/2020/079
(Generalitat Valenciana), and the project i-COOPB20462 (CSIC). G.J.O
thanks the Unidade Acad\^{e}mica de F\'{\i}sica of UFCG (Brazil) and IAFE
(Argentina) for their hospitality during the elaboration of this
work. V.I.A and R.F. thank the Theoretical Physics Department \&
IFIC of the University of Valencia for kind hospitality during the
visits related to this work.

\appendix
\section{Born-Infeld equation derivation} \label{apndxA}

To obtain the antisymmetric part of $\hat q^{-1}$ we first write the
relation (in matrix form) \ben\label{ddd} \hat q^{-1}=( \hat\GG + \b
\hat F)^{-1}=(\hat I + \b\, \hat\GG^{-1}\hat F)^{-1} \hat\GG^{-1}=
\(\sum_{n=0}^{\infty} (-\b)^n (\hat\GG^{-1}\hat F)^n\)\hat\GG^{-1}
\een Noting that $\hat\GG^{-1}\hat F= \GG^{\m\a} F_{\a\m} = \tilde
F^{\m}_{\ \n}$, and using the properties \eqref{SP} and
\eqref{SPrels} of antisymmetric tensors, it is straightforward to
show that the series expansion above can  only have four kinds of
terms, namely, \ben\label{Aq3} \sum_{n=0}^{\infty} (-\b)^n
(\hat\GG^{-1}\hat F)^n = \tilde H \(\d^{\m}_{\ \b}-\b \tilde
F^{\m}_{\ \b} + \b^2 \dual\tilde F^{\m}_{\ \a} \dual\tilde F^{\a}_{\
\b}+\b^3 \tilde P \dual\tilde F^{\m}_{\ \b} \)\ , \een where the
overall scalar factor $\tilde H$ can be directly identified by
solving the identity \ben\label{Aq4} \hat I=( \hat I + \b
\hat\GG^{-1}\hat F)^{-1} ( \hat I + \b \hat\GG^{-1}\hat F) =\tilde H
\(\d^{\m}_{\b}-\b \tilde F^{\m}_{\ \b} + \b^2 \dual\tilde F^{\m}_{\
\a} \dual\tilde F^{\a}_{\ \b}+\b^3 \tilde P \dual\tilde F^{\m}_{\
\b} \)( \d^{\b}_{\n}+ \b \tilde F^{\b}_{\ \n})
 &=&\d^{\m}_{\n}\ ,
\een from which we get $\tilde H =\(1+ \b^{2} 2\tilde S-\b^{4}
\tilde P^2\)^{-1}.$ Thus, the inverse of $q_{\m\n}$ in terms of the
inverse of $\GG_{\m\n}$ reads \ben\label{ddd1} q^{\m\n}&=&
\frac{\tilde\GG^{\m\n}-\b {\tilde F}^{\m\n}+\b^2  \dual {\tilde
F}^{\m}_{\ \ \a} \dual\tilde F^{\a\n}+\b^3 \tilde P \dual\tilde
F^{\m\n}} {1+ \b^{2} 2\tilde S-\b^{4} \tilde P^2} \ . \een

Extracting the antisymmetric part of \eqref{ddd1} and using that
$\dual {\tilde F}^{[\m}_{\ \ \a} \dual\tilde F^{\n]\a}=0$, and that
$\GG_{\m\n}$ is symmetric and non-singular by definition and
therefore $\tilde\GG^{[\m\n]}=0$, we get \ben\label{Aq6} \left( \hat
q^{-1}\right)^{[\m\n]}&=&-\b \frac{{\tilde F}^{\m\n} -\b^2 \tilde P
\dual\tilde F^{\m\n}}{1+ \b^{2} 2\tilde S-\b^{4} \tilde P^2} \ .
\een

On the other hand, writing $q_{\m\n}$ as $q_{\m\n}
=\GG_{\m\r}(\d^{\r}_{\ \n}+ \b \tilde F^{\r}_{\ \n})\,,$ its squared
root determinant takes the form \ben\label{Aq0}
\sqrt{-q}&=&\sqrt{-\GG}\sqrt{\det(\d^{\r}_{\ \n}+ \b \tilde
F^{\r}_{\ \n})}= \sqrt{-\GG}\sqrt{1+2 \b^{2} \tilde S-\b^{4} \tilde
P^2} \ , \een So, using \eqref{FFtilde}, we can finally write down
\ben\label{Aq7} \sqrt{-q} \left[\hat q^{-1}\right]^{[\m\n]} =
-\b\sqrt{-\GG} \frac{{\tilde F}^{\m\n} -b^{-2} \tilde P \dual\tilde
F^{\m\n}}{\sqrt{1+2 b^{-2}\tilde S-b^{-4}\tilde P^2}}
 = -\b\sqrt{-\GG} \wt\FF^{\m\n} \ ,
\een which proves the equation \eqref{VLCBUE}.


\end{document}